\documentclass[12pt]{article}
\usepackage{amsmath}
\usepackage{graphicx}
\usepackage{enumerate}
\usepackage{natbib}
\usepackage{url}
\usepackage{latexsym}
\usepackage{amsthm}
\usepackage{amssymb,amsfonts,bm,amsbsy,bbm}
\usepackage{epsfig}
\usepackage{rotating,lscape}
\usepackage{hyperref}
\usepackage{verbatim}
\usepackage{multirow,color}
\usepackage{setspace}
\usepackage{subfigure}
\usepackage{bbm}
\usepackage{booktabs,caption}
\usepackage[flushleft]{threeparttable}
\usepackage{algorithm}
\usepackage{algorithmic}
\usepackage[algo2e]{algorithm2e}
\usepackage{mathrsfs}  
\usepackage{xcolor}
\usepackage{listings}

\newtheorem{Th}{Theorem}

\newtheorem{Rem}{Remark}

\def\X{{\bf X}}
\def\x{{\bf x}}

\def\Z{{\bf Z}}
\def\z{{\bf z}}
\def\S{{\bf S}}

\def\f{{\bf f}}
\def\Dscr{\mathscr{D}}
\def\mT{\mathcal{T}}
\def\mS{\mathcal{S}}
\def\mI{\mathcal{I}}

\def\mL{\mathcal{L}}
\def\submT{_{\scriptscriptstyle \mT}}
\def\submS{_{\scriptscriptstyle \mS}}
\def\submTL{_{\scriptscriptstyle \mT,L}}
\def\submSL{_{\scriptscriptstyle \mS,L}}
\def\submTU{_{\scriptscriptstyle \mT,U}}
\def\submSU{_{\scriptscriptstyle \mS,U}}
\def\subiota{_{\scriptscriptstyle \iota}}
\def\subiotaU{_{\scriptscriptstyle \iota,U}}
\def\subiotaL{_{\scriptscriptstyle \iota,L}}
\def\supmT{^{\scriptscriptstyle (\mT)}}

\def\supmS{^{\scriptscriptstyle (\mS)}}
\def\supiota{^{\scriptscriptstyle (\iota)}}
\def\supmeta{^{\scriptscriptstyle (\sf meta)}}
\def\supSTRIFLE{^{\scriptscriptstyle (\sf STRIFLE)}}
\def\supmTk{^{\scriptscriptstyle (\mT, k)}}
\def\supmetak{^{\scriptscriptstyle (\sf meta, k)}}

\def\supk{^{\scriptscriptstyle (k)}}
\def\supnegk{^{\scriptscriptstyle (-k)}}
\def\supone{^{\scriptscriptstyle (1)}}
\def\suptwo{^{\scriptscriptstyle (2)}}

\def\ET{E\submT}
\def\ES{E\submS}

\def\bb{{\boldsymbol\beta}}

\def\bz{{\boldsymbol\zeta}}

\def\0{{\bf 0}}
\def\pr{\hbox{pr}}
\def\wh{\widehat}

\newtheorem{Ass}{Assumption}

\def\boxit#1{\vbox{\hrule\hbox{\vrule\kern6pt  \vbox{\kern6pt#1\kern6pt}\kern6pt\vrule}\hrule}}
\def\bse{\begin{eqnarray*}}
	\def\ese{\end{eqnarray*}}
\def\be{\begin{eqnarray}}
\def\ee{\end{eqnarray}}
\def\bsq{\begin{equation*}}
	\def\esq{\end{equation*}}
\def\bq{\begin{equation}}
\def\eq{\end{equation}}

\def\wh{\widehat}

\def\mR{\mathbb{R}}

\def\trans{^{\scriptscriptstyle \sf T}}

\def\bdel{{\boldsymbol\delta}}
\def\bDel{{\boldsymbol\Delta}}

\def\f{{\bf f}}
\def\h{{\bf h}}

\def\I{{\bf I}}

\def\bS{{\bf S}}

\def\v{{\bf v}}
\def\W{{\bf W}}

\def\X{{\bf X}}

\def\x{{\bf x}}

\def\Z{{\bf Z}}
\def\z{{\bf z}}

\def\bSig{{\bf \Sigma}}

\def\log{\hbox{log}}

\def\squarebox#1{\hbox to #1{\hfill\vbox to #1{\vfill}}}
\def\btheta{{\boldsymbol \theta}}

\def\bthetahat{\widehat{\boldsymbol \theta}}

\def\0{{\bf 0}}

\def\pr{\hbox{pr}}
\def\wh{\widehat}
\def\wt{\widetilde}

\def\log{\hbox{log}}

\def\logit{{\mbox{logit}}}

\def\bbhat{\wh\bb}
\def\what{\wh w}
\def\mhat{\widehat{m}}
\def\mbar{\bar{m}}
\def\bthetabar{\bar{\btheta}}

\def\subcor{_{\sf\scriptscriptstyle cor}}
\def\submis{_{\sf\scriptscriptstyle mis}}

\begin{document}
\def\spacingset#1{\renewcommand{\baselinestretch}%
{#1}\small\normalsize} \spacingset{1}

\title{\bf Semi-supervised Triply Robust Inductive Transfer Learning}
\author{Tianxi Cai$^{1,2*}$, Mengyan Li$^{3*}$, Molei Liu$^{4}$\footnote{Authors are listed in alphabetical order. Contact: mengyanli@bentley.edu}\bigskip \\
\small 
$^1${Department of Biostatistics, Harvard T.H. Chan School of Public Health} \\
\small
$^2${Department of Biomedical Informatics, Harvard Medical School}\\
\small 
$^3${Department of Mathematical Sciences, Bentley University} \\
\small 
$^4${Department of Biostatistics, Columbia University Mailman School of Public Health} \\
}
\date{}  

\maketitle

\begin{abstract}
In this work, we propose a {\bf S}emi-supervised {\bf T}riply {\bf R}obust {\bf I}nductive trans{\bf F}er {\bf LE}arning (STRIFLE) approach, which integrates heterogeneous data from a label-rich source population and a label-scarce target population and utilizes a large amount of unlabeled data simultaneously to improve the learning accuracy in the target population. 
Specifically, we consider a high dimensional covariate shift setting and employ two nuisance models, a density ratio model and an imputation model, to combine transfer learning and surrogate-assisted semi-supervised learning strategies effectively and achieve triple robustness. While the STRIFLE approach assumes the target and source populations to share the same conditional distribution of outcome $Y$ given both the surrogate features $\bS$ and predictors $\X$, it allows the true underlying model of $Y \mid \X$ to differ between the two populations due to the potential covariate shift in $\bS$ and $\X$. Different from double robustness, even if both nuisance models are misspecified or the distribution of $Y\mid \bS,\X$ is not the same between the two populations, the triply robust STRIFLE estimator can still partially use the source population when the shifted source population and the target population share enough similarities. 
Moreover, it is guaranteed to be no worse than the target-only surrogate-assisted semi-supervised estimator with an additional error term from transferability detection.
These desirable properties of our estimator are established theoretically and verified in finite samples via extensive simulation studies. We utilize the STRIFLE estimator to train a Type II diabetes polygenic risk prediction model for the African American target population by transferring knowledge from electronic health records linked genomic data observed in a larger European source population. 
\end{abstract}

\noindent%

\vskip 2mm

\spacingset{1.9} 

\section{Introduction}

\subsection{Background}
A precision medicine approach is vitally important for clinical care \citep{ashley2016towards}.
The reliable and precise personalized prediction of clinical outcomes and treatment responses is fundamental for successful disease management. 
More aggressive treatments should be considered for those unlikely to respond to standard therapy while less toxic options may be considered for patients with a favorable prognosis. 
Advances in genetics and the growing availability of electronic health records (EHR) present increased opportunities to develop accurate personalized risk prediction models. Incorporating such models at the bedside makes precise personalized patient care a clinical reality \citep{hodson2016precision}.

Personalized risk prediction modeling, however, faces several challenges in practice. The first challenge is the heterogeneity and imbalance across sub-populations. 
Current prevalent predictive modeling schemes often exhibit biases when applied to heterogeneous and noisy EHR and genomic data, tending to prioritize overall accuracy at the expense of representing under-represented groups. 
This leads to performance disparities caused by varying data completeness, distribution discrepancies, and coding biases.
For instance, about 79\% of GWAS participants are of European descent.
This imbalance, together with genetic differences across ancestral populations, results in polygenic risk scores being several times more accurate for the European group than others \citep{west2017genomics, ferryman2018fairness, martin2019clinical}. 
The presence of heterogeneity and the relatively low sample size of the under-represented groups highlight the need for robust transfer learning techniques to effectively improve prediction accuracy in the underrepresented sub-populations and achieve fairness in precision medicine.

The second challenge arises from the lack of direct observation of the clinical outcomes needed for training supervised prediction models.
Precise clinical outcome data often require extensive manual chart review, which is not scalable. Alternative sources like diagnosis codes are readily available but lack accuracy. To efficiently leverage both the limited labeled data and the imperfect surrogates, it is critical to incorporate {\em semi-supervised (SS) learning} strategies to combine information from both labeled and unlabeled observations. The third challenge comes from the high dimensionality of the feature space. Both the number of candidate predictors, such as genetic markers, and the number of candidate surrogates from EHR can be large. 

Much progress has been made in transfer learning and SS learning in recent years, both in low and high dimensional settings. However, there is a paucity of literature on robust transfer learning under SS settings to leverage the large amount of unlabeled source and target data while the amount of target labeled data is small. We aim to address this gap by developing a {\bf S}emi-supervised {\bf T}riply {\bf R}obust {\bf I}nductive trans{\bf F}er {\bf LE}arning (STRIFLE) method that effectively leverages such data structure to improve the estimation and prediction accuracy of risk modeling in the underrepresented target population. Specifically, we consider a covariate shift setting, which allows the predictors, $\X$, and the surrogates of the labels, $\S$, to have different distributions in the target and the source population. We assume that the two populations share the same conditional distribution of the outcome $Y$ given $\X$ and $\S$, but allow the conditional distribution of the outcome $Y$ given $\X$ to differ between the two populations. In EHR applications, since $\S$ can be highly predictive of the response, $Y$ given $\X$ and $\S$ tends to be more consistent across populations compared with $Y$ given $\X$.
In both populations, we have a small number of labeled observations and a large number of unlabeled observations, respectively. Further, the number of the candidate predictors and the surrogates can be much larger than the number of labeled observations.

\subsection{Related Literature and Our Contributions}

Our work is closely related to several tracks of recent literature in statistical learning. In this section, we shall introduce them separately and summarize our methodological novelty and contribution in view of the existing literature.

\paragraph{Surrogate-assisted semi-supervised learning.} In recent years, extensive research has been conducted on SS learning that makes use of the large amount of unlabeled data to improve the estimation and prediction accuracy with the small amount of labeled data \citep[e.g.]{zhu2005semi, chakrabortty2018efficient,gronsbell2022efficient}. Moreover, in biomedical and EHR studies, highly predictive surrogates of the disease status, such as the diagnostic codes, medication prescriptions, and laboratory tests are readily available for the unlabeled observations. To improve the learning accuracy, \cite{hou2021surrogate} proposed a robust surrogate-assisted semi-supervised (SAS) method that imputes the unobserved outcome (e.g. the disease status) by utilizing the candidate predictors (e.g. genetic variants) as well as the surrogate EHR features via a sparse working imputation model. 
\cite{zhang2022prior} proposed an SS method that adaptively incorporates prior knowledge from EHR surrogates by shrinking the high dimensional sparse estimator towards a direction derived from the unlabeled data with surrogates. However, none of these existing methods can handle covariate shift between the source population and the target population, which could lead to biased or inefficient estimation of the target parameters. 

\paragraph{Covariate shift correction.} Distributional shift of the covariates between the training (source) and testing (target) data has been frequently studied in recent years. While importance weighting \citep[e.g.]{huang2007correcting}
is not robust to misspecified models or excessive estimation errors of the density ratio model. In recent years, methods have been developed for doubly robust covariate shift correction \citep[e.g.]{liu2020doubly,zhang2021double,zhou2022doubly}. Motivated by semiparametric theory, they introduce an imputation model for $Y$ to augment the importance weighted estimating equations. The resulting estimator is valid when either the density ratio or the imputation model is correctly specified. The main limitation of this strategy is that it purely relies on labels from the source population and fails to give a valid estimation when both models are misspecified.
The limitation is further exacerbated if the conditional model of $Y$ varies between the source and target populations.
Under the scenario considered in our paper, there is a space to overcome this limitation due to the availability of a small amount of labeled data from the target population.

\paragraph{Knowledge transferring regression.}
Recently,  \cite{bastani2021predicting}, \cite{li2020transfer}, \cite{ tian2021transfer}, and \cite{li2023estimation} proposed trans-Lasso, a transfer learning method for high dimensional linear models and generalized linear models (GLM) with Lasso penalty \citep{tibshirani1996regression} to improve the learning accuracy on target population by adaptively borrowing information from potentially transferable source data. 
Their core idea is to shrink source estimators learned using source labeled data towards target parameters through Lasso penalties. Based on this trans-Lasso framework, \cite{li2021targeting} proposed a federated transfer learning approach that integrates heterogeneous data from diverse populations and multiple healthcare institutions. 
Different from us, existing methods on trans-Lasso neither adjust for covariate shift between the source and target data nor incorporate unlabeled target observations with surrogates. As will be verified in our theoretical and numerical studies, this could cause a large efficiency loss in our setting. 

\paragraph{Semi-supervised transfer learning}

Our proposed STRIFLE method also relates to the SS transfer learning literature under a setting where the target data has abundant unlabeled samples to assist the transfer learning, sometimes as well as the source. For example, \cite{abuduweili2021adaptive} proposed an adaptive consistency regularization method for deep learning that involves two penalization terms ensuring adaptive knowledge transfer and representation consistency between the labeled and unlabeled target samples, respectively. \cite{jakubovitz2023information} improved the
transferability of neural networks (NN) by incorporating regularization terms on the large unlabeled target data based on mutual (or Lautum) information. In the context of statistical learning closer to us, \cite{zhang2023double} developed a doubly robust SS method for the mean estimation on a target population with covariate shift. 
\cite{zhang2023semi} further extended this framework to the SS transfer learning of casual effects. However, this track of work did not address certain main challenges in transfer learning, including model shift and high-dimensionality of the target models. Neither did they handle potential misspecification of the parametric nuisance models adequately.

\paragraph{Our contribution.}
The STRIFLE method we propose stands out for being both surrogate-assisted and model-assisted.
Assisted by predictive surrogates, we employ a working and nuisance imputation model under a moment constraint to impute the outcomes for the unlabeled target observations. To integrate the label-rich source population, we introduce a working and nuisance density ratio model embedded in a penalized transfer learning procedure when estimating the imputation model.
The novel embedded density ratio model, estimated using unlabeled observations from the two populations, can potentially increase the transferability of the source population. 

Unlike the naive combination of transfer learning and SS learning, which is vulnerable to model misspecification and may lead to erroneous results practically, STRIFLE unifies transfer learning and SAS techniques using the two nuisance models as bridges, and achieves `triple robustness'.
Building upon the concept of `double robustness' in the literature, where robustness is achieved if at least one of two nuisance models is correctly specified, our approach introduces a third layer of robustness, augmented by a bias correction step and a transferability detection procedure, benefiting from the availability of a small amount of labeled target data. The concept of triple robustness is characterized by three key features. 
First, it extends beyond double robustness by being more resilient to misspecification of nuisance models. 
Even if both nuisance models are misspecified, STRIFLE can still partially utilize source labeled data to boost statistical efficiency, provided there are sufficient similarities between the source population shifted based on the working density ratio model and the target population.
Second, it is robust to  potential violation of the conditional distributions of $Y$ given $\S$ and $\X$ being identical between populations, as long as 
the shifted source imputation model and the target imputation model are similar enough. Third, 
STRIFLE is an adaptive transferring approach in the sense that it is guaranteed to be no worse than the target-only SAS estimator with an additional error term from transferability detection. 
In addition, our framework is not restricted to Lasso penalty, and allows the use of non-convex penalties such as SCAD \citep{fan2001variable} and MCP\citep{zhang2010nearly}.
To the best of our knowledge, this is the first work that leverages information from the source population and simultaneously utilizes unlabeled observations from both populations, achieving a higher level of robustness.

\subsection{Outline of the Paper}

We introduce the problem setup with an overview of the STRIFLE algorithm and present detailed methodology in Section \ref{sec:method}, establish theoretical properties of our STRIFLE estimator in Section \ref{sec:theory}, examine its finite-sample performance through simulation studies in Section \ref{sec:simulation}, and apply the STRIFLE algorithm to develop a polygenic risk prediction model for Type II Diabetes (T2D) in African American (AA) population based on single nucleotide polymorphisms (SNPs) and EHR features with the European population as the source in Section \ref{sec:realdata}. The paper is concluded with some discussions in Section  \ref{sec:dicussion}.

\section{ Methodology}\label{sec:method}
\subsection{Problem Setup}\label{subsec:setup}
For the $i$-th observation, $Y_i \in \mR$ denotes the outcome variable,  $\X_i\in \mR^p$ denotes the $p$-dimensional candidate predictors including the intercept term as its first element, and $\S_i \in \mR^q$ denotes the $q$-dimensional surrogates for $Y_i$. 
We let $\mT$ and $\mS$ denote the target and source populations; $L$ and $U$ index the labeled and unlabeled subsets; and $\mI_{\iota,\jmath}$ denote the index set of observations from target/source labeled/unlabeled populations with $\iota \in \{\mT,\mS\}$ and $\jmath \in \{L,U\}$. We assume that $Y$ is only observed in the labeled subsets while $\Z = (\X\trans, \S\trans)\trans$ is observed in both labeled and unlabeled data. We let 
$\Dscr\submT = \Dscr\submTL \cup \Dscr\submTU$ and $\Dscr\submS = \Dscr\submSL \cup \Dscr\submSU$ denote the observed data from the target and the source populations with 
$\Dscr\subiotaL = \{(Y_i,\Z_i\trans)\trans, i \in \mI\subiotaL\}$, $\Dscr\subiotaU = \{\Z_i, i \in \mI\subiotaU\}$ , 
$\iota = \mT, \mS$.
Let $|\mathcal{I}\submTL| = n\submT$, $|\mathcal{I}\submTU| = N\submT$, $|\mathcal{I}_{\mS, L}| = n\submS$,  $|\mathcal{I}_{\mS, U}| = N\submS$, $d = p+q$, and $n = n\submS +n\submT$.
We consider the high dimensional setting where $p$ and $q$ grow with $n\submT$ and can be much larger than $n\submT$ and $n$, and allow $n\submS$ and $n\submT$ to vary freely. While this work is primarily motivated by an SS context, our theoretical results also accommodate scenarios where 
$N\submT \leq n\submT$.

Let $p\supiota_\Z$, $p\supiota_{Y\mid \Z}$ , $p\supiota_{\S\mid \X} $, and $p\supiota_{Y\mid \X}$  denote the probability density of $\Z_i$, the conditional density of $Y_i$ given $\Z_i$,  the conditional density of $\S_i$ given $\X_i$, and the conditional density of $Y_i$ given $\X_i$, respectively, in the population $\iota$, for $\iota \in \{\mT,\mS\}$. 
We consider covariate shift structure in this work, that is, we assume $p\supmS_{Y\mid \Z}(y, \z) = p\supmT_{Y\mid \Z}(y, \z)$
while allowing  $p\supmS_\Z(\z)$ to be different from $p\supmT_\Z(\z)$.
Specifically, 	
we allow $p\supmS_{\S\mid \X} \neq p\supmT_{\S\mid \X}$ and consequently, $p\supmS_{Y\mid \X}$ can be different from  $p\supmT_{Y\mid \X}$. 
We let  $E\subiota(\cdot)$ denote the expectation taken over the distribution in the population $\iota \in \{\mS,\mT\}$.

We aim to develop a prediction model for $Y\mid \X$ in $\mT$ under a {\em working model} $\ET(Y_i \mid \X_i)=g(\X_i\trans\bb)$, where $g(\cdot)$ is a known inverse link function. Specifically, our goal is to estimate $\bb_0 \in \mR^p$, the solution to 
\be
\label{eq:beta_MomentCondition}
\ET[\X_i \{Y_i - g(\X_i\trans\bb)\}]=\0.
\ee
We allow $g(\X_i\trans\bb)$ to be misspecified for $\ET(Y_i \mid \X_i)$ and  
allow $\bb_0$  to be dense.
Although $\S_i$ can be strongly informative of $Y_i$,  the outcome prediction model in (\ref{eq:beta_MomentCondition}) does not include $\S_i$.
In the motivating example of developing a polygenic risk prediction model for T2D,  $Y$ is the T2D status, $\S$ comprises EHR features like Hemoglobin A1C (HbA1c) test mentions and metformin prescriptions, and $\X$ includes demographic variables and genetic markers. 
Though the surrogates represent noisy proxies of the disease status and can be used to impute the missing $Y$, they are not suitable as predictors when the prediction is performed at some baseline time prior to the ascertainment of $Y$ or its noisy proxy $\S$.

\subsection{Overview of the STRIFLE Algorithm}\label{subsec:alg}
We first give an overview of the STRIFLE procedure. To leverage the unlabeled data and the surrogates, STRIFLE employs a series of robust imputation strategies under a {\em working} imputation model for $Y \mid \Z$:
$m(\Z) = \ET(Y \mid \Z) = \ES(Y \mid\Z) = g(\Z\trans\btheta)$.
For any fitted imputation model $\mhat(\Z) = g(\Z\trans \bthetahat)$, let $\bthetabar$ be the limit of $\bthetahat$ when the sample size goes to infinity, and $\mbar(\Z) = g(\Z\trans \bar\btheta)$ be the limiting imputation model.
We construct an SS estimator for $\bb_0$ utilizing the fitted imputation model as
$\wh \bb (\bthetahat) = \arg \min_{ \bb \in \Omega_\bb} \mL_{N\submT+n\submT}^{\bb}(\bb; \bthetahat) + p_{\lambda_{\bb}}(\bb)$,
where 
\be
\label{eq:lossbb}
\mL_{N\submT+n\submT}^{\bb}(\bb;\bthetahat) = -\frac{1}{N\submT + n\submT} \left[\sum_{i\in \mathcal{I}\submTU } Q\{g(\x_i\trans \bb), g(\z_i\trans\bthetahat)\} + \sum_{i\in  \mathcal{I}\submTL} Q\{g(\x_i\trans \bb), y_i\}\right],
\ee
$Q(\cdot, \cdot)$ is the  log-(quasi-)likelihood function (after dropping any dispersion parameter), i.e., 
$Q\{g(\x_i\trans\bb),\widehat m(\z_i)\} = \widehat m(\z_i) \x_i\trans \bb - G(\x_i\trans \bb)$,  $Q\{g(\x_i\trans\bb), y_i\} = y_i \x_i\trans \bb - G(\x_i\trans \bb)$, 
$G(u) = \int_0^u g(u^\prime) du^\prime$,  $p_{\lambda_{\bb}}(\bb)$ is a penalty function on $\bb$ with a tuning parameter $\lambda_\bb$, and $\Omega_\bb$ is the $p$-dimensional parameter space. Detailed definition of $\Omega_\bb$ is given in Section \ref{sec:theory}. 

By the definition of $\bb_0$ in (\ref{eq:beta_MomentCondition}), it is straightforward to show that $\wh\bb$ is consistent for $\bb_0$ provided that  
\begin{equation}
\label{eq:impute}
E\submT[\X_i \{ Y_i - \mbar(\Z_i)\}] = \0.    
\end{equation}
A straightforward way to ensure the consistency of $\wh\bb$
is to solely rely on the target labeled data for estimating $\btheta$ as in \cite{hou2021surrogate}.
However, this approach misses the opportunity to benefit from the abundant labeled data in $\mS$.
In covariate shift setting where $p_{Y\mid \Z}^{(\mS)} = p_{Y\mid \Z}\supmT$ but $p_{Y\mid \X}^{(\mS)}$ can
differ significantly from $p_{Y\mid \X}\supmT$, the learning accuracy of $\bb_0$ can be improved if we can borrow useful information from $\mS$ when estimating $m(\z)$.
Thus, our STRIFLE algorithm centers its strategy on robust and efficient estimation of $\btheta$ by leveraging useful information from $\mS$. 

Specifically, we employ a {\em working} density ratio model to improve the transferability of the source data in a double-robust manner.
Note that when both {\em working} imputation and density ratio models are misspecified, the estimation of $\btheta$ can be inconsistent.
To address this, we use the limited labeled target data for bias correction. However, if the {\em working} models are severely flawed, $n\submT$ may be insufficient for full bias correction.
To avoid negative transfer, we include a transferability detection step. 
If the discrepancy between the shifted source and target imputation models is too large, we opt to not use the source data.
The STRIFLE estimation of $\btheta$ consists of three key components: 
\begin{enumerate}
	\item fitting a  {\em working} density ratio model to estimate $w(\z) =p\supmT_\Z(\z)/ p\supmS_\Z(\z)$ as $\what(\z)$; 
	\item constructing an initial target-only  estimator of the imputation parameter $\btheta$ as $\bthetahat\supmT$ as well as a
	meta-learning estimator, $\bthetahat\supmeta$, using $\Dscr\submT\cup \Dscr\submS$ and $\what(\z)$; 
	\item constructing the STRIFLE estimator of $\btheta$, $\bthetahat\supSTRIFLE$, as a convex combination of $\bthetahat\supmT$ and $\bthetahat\supmeta$ to overcome potential negative transfer.
\end{enumerate}
The final STRIFLE estimator for $\bb$ is then obtained as
$\wh\bb\supSTRIFLE = \wh\bb(\bthetahat\supSTRIFLE)$.
We detail each of the components below.

\subsection{Estimation of the Density Ratio Model}\label{subsec:w}
To potentially increase the transferability of the source population and efficiently leverage the source data, we fit a {\em working} density ratio model utilizing the large number of unlabeled observations from both populations for 
$p\supmT_\Z(\z)/ p\supmS_\Z(\z)$:
\begin{equation}
	w(\z) = \exp\{ \f(\z)\trans\bz\} , \label{model-DR}
\end{equation}
where $\f(\cdot) \in \mathbb{R}^L$ is a vector of basis functions that can be used to capture non-linear effects. 
Equation (\ref{model-DR}) encapsulates a wide range of models. $\f(\cdot)$ can be derived from prior knowledge related to covariate shifts or estimated directly from unlabeled data. 
The methodology with estimated basis functions $\wh\f(\z)$ aligns well with various machine learning (ML) techniques.
For example, in multi-layer NN, the final layer acts as a simple linear classifier \citep{papyan2020prevalence}, 
and $\wh\f(\z)$ can be obtained by fitting the NN (excluding the last layer) with the unlabeled data. 
However, it's important to recognize that even when employing ML methods for density ratio estimation, the challenges of model misspecification and excessive error persist, as discussed in \cite{dukes2021doubly}. Hence, the bias correction and the transferability detection given in Section \ref{subsec:Trans_w} and \ref{subsec:negativeTrans} are still necessary. This issue largely hinges on the potential non-convexity of ML methods
, complicated tuning procedures, and the balance between model complexity and the size of unlabeled data.
In Section \ref{sec:theory}, the theoretical results are derived assuming the known nature of $\f(\cdot)$. Further discussions on $\widehat \f(\z)$ and ML methods are presented in Remark \ref{rem:density}.

One key observation related to estimating $\bz$ is that 
if $w(\z)$ is correctly specified, for any measurable function $\h(\z)$, we have $\ES\{w(\Z_i) \h(\Z_i)\} = \ET\{\h(\Z_i)\}$.
Here we let $\h(\z) = \f(\z)$, as also adopted in  \cite{tan2020model} and 
\cite{imai2014covariate}. 
Other $\h(\z)$ can be used to construct the loss function for $\bz$ as discussed in Section S.1 of Supplementary Materials.
We define the population parameter $\bz_0$ as the minimizer of 
$E\submS\left[\exp\left\{\f(\Z_i)\trans \bz\right\}\right]- E_{\mT}\left\{\f(\Z_i)\trans\bz\right\}$.
Many standard estimators can be used to estimate $\bz_0$.
In our theoretical derivations, we do not restrict to any specific estimator provided that its convergence rates satisfy Assumption \ref{ass:w(z)} given in Section \ref{sec:theory}. 

One such estimator for $\bz_0$ can be constructed as $\wh \bz = \arg \min_{\bz \in \Omega_{\bz}} \mL_{N}^w(\bz)  +  p_{ \lambda_\bz}(\bz)$,
where
$\mL_{N}^w(\bz) =N_{\mS}^{-1} \sum_{i\in  \mathcal{I}_{\mS, U}}\exp\{\f(\z_i)\trans\bz\} - N\submT^{-1} \sum_{i\in \mathcal{I}\submTU} \f(\z_i)\trans\bz$, $p_{ \lambda_\bz}(\bz)$ is a penalty function with a tuning parameter $\lambda_\bz$,  and $\Omega_\bz$ is the feasible parameter space of $\bz$. 
One may choose $p_{ \lambda_\bz}(\bz)$ as commonly used penalty functions including the ridge, Lasso, SCAD, etc. 
Here and after, with a slight abuse of notation, we use $p_\lambda(\cdot)$ to denote different types of penalty functions with a tuning parameter $\lambda$.
We show in Section S.2 of Supplementary Materials that $\wh \bz$ obtained with a Lasso penalty and $\Omega_\bz = \mathbb{R}^L$ satisfies Assumption \ref{ass:w(z)}.

\subsection{Target-only and Meta-learning of the Imputation Model}\label{subsec:Trans_w}
We next construct two estimators for $\btheta$ in $m(\z)$, one using $\Dscr\submTL$ only and one via meta-learning combining $\Dscr\submT$ and $\Dscr\submS$. These estimators are derived to achieve both robustness and efficiency of the estimation of the target parameter $\bb_0$ in the presence of potential model misspecifications. To this end, we first provide two key observations about the estimation of  $m(\z)$. 
First, the target unlabeled observations can be utilized to consistently estimate $\bb_0$ with missing labels imputed by any consistent estimator of $\mbar(\z)$ satisfying the moment condition (\ref{eq:impute}).
We define 
\be 
\label{eq:theta0}
\bthetabar = \arg\max_{\btheta } E\submT\left[	Q\left\{g(\Z_i\trans\btheta), Y_i \right\} \right],
\ee
where $Q\{g(\z_i\trans \btheta), y_i\} = y_i\z_i\trans \btheta - G(\z_i\trans \btheta)$,
and the resulting $\mbar(\z)=g(\z\trans\bthetabar)$ satisfies  the moment condition (\ref{eq:impute}), even if the imputation model is misspecified.
With target labeled data, one straightforward estimator of $\bthetabar$ is 
$\bthetahat\supmT = \arg\min_{\btheta \in \Omega_{\btheta}\supmT}
\left[-{n\submT}^{-1}\sum_{i\in \mathcal{I}\submTL}  
Q\left\{ g(\z_i\trans \btheta), y_i\right\}  + p_{\lambda}(\btheta)\right]$,
where $p_{\lambda}(\btheta)$ is a penalty function on $\btheta$ with a tuning parameter $\lambda$, and $\Omega_\btheta\supmT$ is the feasible parameter space of $\btheta$. Under standard regularity conditions in high dimensional settings,  $\bthetahat\supmT $ converges to $\bthetabar$. 
Using $\bthetahat\supmT$ in (\ref{eq:lossbb}) leads to SAS estimator for $\bb_0$.

The second key observation is that, under our covariate shift setting and the moment condition (\ref{eq:impute}),  when the imputation model or the density ratio model is correctly specified,  we have
$E\submS[w_0(\Z_i) \Z_i  \{ Y_i -\bar m(\Z_i) \}] = \0$.
We further define
\begin{align}
	\label{eq:theta*}
	\btheta^* =\arg\max_{\btheta}  \left\{\frac{n\submS}{n}E\submS \left[w_0(\Z_i) Q\left\{ g(\Z_i\trans\btheta), Y_i\right\} \right] + \frac{n\submT}{n}E\submT \left[ Q\left\{ g(\Z_i\trans\btheta), Y_i\right\}\right]\right\},
\end{align}
and $m^*(\z) = g(\z\trans\btheta^*)$. 
When either $m(\z)$ or $w(\z)$ is correctly specified, $\btheta^* = \bthetabar$ and one may estimate $\bthetabar$ by maximizing an empirical version of (\ref{eq:theta*}) to derive a doubly robust estimator for $\bthetabar$. 
However, when both models are misspecified or $p_{Y|\Z}$ differs between populations, $\btheta^* \neq \bthetabar$ and such an approach could lead to a biased estimator for $\bthetabar$.

Inspired by the idea of oracle Trans-Lasso \citep{li2020transfer, tian2021transfer}, 
we propose a two-step density-ratio-embedded {\em meta-learning} algorithm to estimate $\bthetabar$ by correcting the potential bias $\bdel_0 = \bthetabar - \btheta^*$, as summarized in Algorithm \ref{alg}.
Different from oracle Trans-Lasso, we deal with a more complex loss function with an estimated {\em working} density ratio embedded in it.

\begin{algorithm}[tbh]
	\singlespacing
	\caption{Robust Meta-Learning of $\bthetabar$}
	\begin{algorithmic}\label{alg}
		\STATE  1.  Using $\Dscr\submSL\cup \Dscr\submTL$ along with $\wh w(\z) = \exp(\z\trans \widehat\bz)$, we estimate $\btheta^*$ as $\wt \btheta  = \arg \min_{\btheta \in \Omega_\btheta}\mL_n(\btheta; \wh \bz) + p_{\lambda_{\btheta}}(\btheta)$,
		where 
		\begin{equation}
			\label{eq:losstheta*}
			\mL_n(\btheta; \bz)
			=- \frac{1}{n}\sum_{i\in \mathcal{I}\submSL\cup \mathcal{I}\submTL}w(\z_i; \bz)^{I(i\in \mathcal{I}\submSL)}Q\left\{g(\btheta\trans\z_i), y_i \right\},
		\end{equation}
		$I(i\in \mathcal{I}_{\mS, L})$ is an indicator function, $p_{\lambda_{\btheta}}(\btheta)$
		is a penalty function on $\btheta$ with a tuning parameter $\lambda_\btheta$, and $\Omega_\btheta$ is the feasible parameter space with the detailed definition given in Section \ref{sec:theory}.
		
		~\\
		
		\STATE 2. Estimate $\bdel_0$ using  $\Dscr\submTL$ with $\wt \btheta$ as
		$\wh \bdel = \arg\min_{\bdel \in \Omega_{\bdel}}
		\mL_{n\submT}\supmT(\wt \btheta + \bdel)  + p_{\lambda_{\bdel}} (\bdel)$,
		to construct the meta-learning estimator $\bthetahat\supmeta = \wt \btheta + \wh \bdel$, where
		\begin{equation}
			\mL_{n\submT}\supmT(\wt \btheta + \bdel) :=  -\frac{1}{n\submT} \sum_{i\in \mathcal{I}\submTL} Q[g\{ \z_i\trans (\wt \btheta+  \bdel)\},y_i], 
		\end{equation}
        $p_{\lambda_{\bdel}}(\bdel)$ is a penalty function on $\bdel$ with a tuning parameter $\lambda_\bdel$, and $\Omega_\bdel$ is the feasible parameter space with the detailed definition given in Section \ref{sec:theory}.
		
		{\footnotesize *	With a slight abuse of notation, $p_{\lambda_\btheta}$ and $p_{\lambda_{\bdel}}$ can be different penalty functions with tuning parameter $\lambda_\btheta$ and $\lambda_{\bdel}$, respectively.
		}
	\end{algorithmic}
\end{algorithm}

In the first step, we estimate $\btheta^*$ by minimizing the empirical version of negative of (\ref{eq:theta*}) with a penalty function. 
Under covariate shift setting, 
when at least one of the two working models is correctly specified, $\bdel_0 = \0$ and $\wt \btheta$ is consistent for  $\bthetabar$ under standard high dimensional conditions. In this case, source labeled data can be fully utilized. 
When $\btheta^*\neq \bthetabar$ yet $\bdel_0$ is relatively close to zero,  
with the second bias correction step, 
the source labeled data can still be partially utilized and hence the meta-learning estimator $\bthetahat\supmeta$ can still outperform the target-only estimator $ \bthetahat\supmT$. 
However, the second step is unable to fully correct the bias in $\wt \btheta$ when $\bdel_0$ is too dense to be fully recovered with limited labeled target data. This negative transfer phenomenon is verified in both our theoretical analyses in Section \ref{sec:theory} and numerical studies in Section \ref{sec:simulation}. We overcome this in
our final step of the STRIFLE estimation as
detailed below.

\subsection{STRIFLE Estimator of the Imputation Model}\label{subsec:negativeTrans}
To overcome negative transfer, we define our STRIFLE estimator for $\bthetabar$ as a convex combination of  $\bthetahat\supmT$ and $\bthetahat\supmeta$, and estimate the combination weight via data splitting. Specifically, we randomly split data in $\mathcal{I}\submTL$ into two halves, $\mathcal{I}\submTL\supone$  and $\mathcal{I}\submTL\suptwo$.
We use data in $\mathcal{I}\submTL\supk$ to construct the target-only estimator $\bthetahat\supmTk$, and use data in $\mathcal{I}\submSL$ and $\mathcal{I}\submTL\supk$ to construct the meta-learning estimator $\bthetahat\supmetak$ based on Algorithm \ref{alg}. The remaining data in $\mathcal{I}\submTL\setminus\mathcal{I}\submTL\supk$ are used to estimate the coefficient in the convex combination. Inspired by \cite{tian2021transfer}, 
we consider a binary coefficient $\rho \in\{0,1\}$, inferring whether $\mS$ is transferable, which is estimated by comparing the loss $\mL^{\rho\supnegk}_{n\submT/2} ( \btheta )  = 2/n\submT \sum_{i \in \mathcal{I}\submTL\setminus\mathcal{I}\submTL\supk} G(\z_i\trans\btheta) -y_i \z_i\trans \btheta$ at $\bthetahat\supmTk$ and $\bthetahat\supmetak$. To mitigate the efficiency loss of data splitting, cross-fitting can be employed and
\begin{equation}
	\label{eq:LM_ineq}
	\wh \rho = I\left\{\frac{1}{2}\sum_{k=1}^2 \mL^{\rho\supnegk}_{n\submT/2} (\bthetahat\supmetak)
	\leq \frac{1}{2}\sum_{k=1}^2\mL^{\rho\supnegk}_{n\submT/2} (\bthetahat\supmTk) - \epsilon_0 \right\}, 
\end{equation}
where $\epsilon_0$ is a non-negative constant chosen to achieve theoretical guarantee.
See  Section S.4
of  Supplementary Materials for more details of the choice of $\epsilon_0$. 

The final STRIFLE estimator for $\bthetabar$ is defined as 
$\bthetahat\supSTRIFLE = \wh\rho \bthetahat\supmeta + (1 - \wh\rho)\bthetahat\supmT$, 
which is no worse than $\bthetahat\supmT$ with an additional error from estimating $\rho$ even if $\bdel_0$ significantly deviates from the zero vector. Consequently, $\wh\bb\supSTRIFLE $  is no worse than the target-only SAS estimator accompanied by an extra error resulting from transferability detection. When $\bdel_0$ is equal to or close to zero
and $n\submS$ is much larger than $n\submT$, $\wh\bb\supSTRIFLE$ can be substantially more efficient than SAS estimator.

\section{Theoretical Results} \label{sec:theory}
We start by introducing some notations. For a vector $\v = (v_1, \ldots, v_p)\trans \in \mathbb{R}^p$, we
define $\|\v\|_0 = |\rm{supp}(\v)|$, where $\rm{supp}(\v) = \{j: v_j
\neq 0\}$
and $|A|$ is the cardinality of a set $A$. For $S \subseteq \{1, \ldots, p\}$, let $\v_S = \{v_j: j \in S\}$ and $S^c$ be the complement of $S$ .
Denote $\|\v\|_{\infty} = \max_{1\leq j \leq p}|v_j|$, and $\|\v\|_r = \sum_{j=1}^p |v_j|^r$, for $r\in (0, 1]$.
Define $\mathbb{B}_r^p(R) = \{\v \in \mathbb{R}^p: \|\v\|_r \leq R\}$, for $r\in[0,1]$.
For two positive sequences $a_n$ and $b_n$, we use $a_n \precsim b_n$
to denote $a_n \leq Cb_n$ for some constant $C>0$,
and use $a_n \asymp b_n$ to denote $C\leq a_n/b_n \leq C^{\prime}$ for
some constants $C, C^{\prime}>0$.

We then introduce an assumption on the convergence rates of density ratio estimation, essential for the theoretical basis of our STRIFLE estimator. 
\begin{Ass}
\label{ass:w(z)}
Denote by $\Omega_0$ the event that 
$\|\wh \bz - \bz_0\|_1 \leq C_{\bz}$, and $\|\wh \bz - \bz_0\|_2^2 = o(1)$,
where $C_{\bz}$ is a positive constant, and $\pr(\Omega_0)$ goes to 1 when the sample size goes to infinity. 
\end{Ass}

\begin{Rem}\label{rem:density}
As mentioned in Section \ref{subsec:w}, there is great flexibility in estimating the density ratio model. Here, we focus on the parametric model $\exp\{\f(\z)\trans\bz\}$ with pre-specified $\f(\cdot)$, which is allowed to be misspecified, to simplify theoretical derivations.
As detailed in Section S.1
of Supplementary Materials, the density ratio estimation can be reformulated as a binary classification task, enabling the use of ML techniques. 
In this case, Assumption \ref{ass:w(z)} can be modified to
`$\ES[\{\widehat w(\Z) - w_0(\Z)\}^2] = o(1)$, and $\sup_\Z w_0(\Z)/\widehat w(\Z)$ is bounded with probability goes to 1 when the sample size goes to infinity.'
 Given sufficient unlabeled data from both populations, recent advances in convergence rates of ML methods, such as deep NN \citep{farrell2020deep, farrell2021deep} and generalized random forests \citep{athey2019generalized}, ensure that these methods do not compromise the theoretical performance of our STRIFLE estimator under corresponding conditions. 
\end{Rem}

Regarding sparsity assumptions, similar to \cite{li2020transfer}, we allow $\bdel_0$ to be soft sparse.  
For $\bar\btheta$, we consider two sparsity scenarios. 
In scenario 1, like existing work on TransLasso \citep{li2020transfer, tian2021transfer, li2021targeting, li2023estimation}, we assume $\bar\btheta$ is $l_0$ sparse.
While in scenario 2, we assume $\bar\btheta$ and $\bdel_0$ are both $l_r$ sparse, and consequently, $\btheta^* = \bar \btheta - \bdel_0$ is also $l_r$ sparse for a fixed $r \in [0,1]$. 
In this section, we only present the detailed theoretical results for 
$l_0$ sparse $\bar\btheta$ and $l_1$ sparse $\bdel_0$, driven by the simplicity and ease of interpretation inherent in this specific scenario.
We provide a brief explanation of the theoretical findings in other scenarios in Remark \ref{NT_sc2}. The detailed theoretical results in general settings are deferred to Sections S.5-S.8 of Supplementary Materials.

We defer other Assumptions S1-S5 on the design matrices, models, penalty functions, sparsity of $\bdel_0$, $\bthetabar$ and $\bb_0$ to Section S.3
of Supplementary Materials.
In short, in the scenario where $\bar\btheta$ is $l_0$ sparse and $\bdel_0$ is $l_1$ sparse, 
we assume 
$\bdel_0 \in \mathbb{B}_1^d(R_{\bdel, 1})$,
$R_{\bdel, 1}\precsim \sqrt{n\submT/\log~d}$, 
$\bthetabar \in \mathbb{B}_r^d(R_{\bar\btheta, r})$, $\bb_0 \in \mathbb{B}^p_r(R_{\bb, r})$, $R_{\bar\btheta, r}\precsim (n\submT/\log~d)^{1-r/2}$, and
$R_{\bb, r}\precsim \{( N\submT+n\submT)/\log~d\}^{1-r/2}$
for $r=0, 1$. With a slight abuse of notation, we define the true supports of $\bthetabar$ and $\bb_0$ as $S_\btheta$ and $S_\bb$ with cardinality $s_\btheta$ and $s_\bb$, respectively.
We also assume  $\lambda_\btheta \asymp \sqrt{n^{-1}\log ~d}$, $\lambda_\bdel \asymp\sqrt{n\submT^{-1}\log ~d}$, and 
$\lambda_\bb  \asymp\sqrt{(N\submT +n\submT)^{-1}\log~p}$.
We establish theoretical results under the framework of \cite{negahban2012unified} and \cite{loh2015regularized} allowing for nonconvex penalties like SCAD and MCP.
To facilitate theoretical derivations, similar to \cite{loh2015regularized}, we define the feasible parameter spaces as 
$\Omega_\btheta = \{\btheta \in \mathbb{R}^d : \|\btheta\|_1 \leq R_{\bar\btheta, 1}\}$, 
$\Omega_\bdel = \{\bdel \in \mathbb{R}^d : \|\bdel\|_1 \leq R_{\bdel, 1}\}$, and
$\Omega_\bb = \{\bb \in \mathbb{R}^p : \|\bb\|_1 \leq R_{\bb, 1}\}$. Note that if the loss functions with penalties  are convex, the feasible parameter spaces extend to 
$\Omega_\btheta = \Omega_\bdel = \mathbb{R}^d$, and $\Omega_\bb = \mathbb{R}^p$.

\subsection[]{ Theoretical Properties of  $\bthetahat\supSTRIFLE$}
\label{subsection: theta}

We first establish the error bounds of $\bthetahat\supmeta$ in the following Theorem with proofs given in Section S.6 of Supplementary Materials.

\begin{Th}\label{th:1}
Suppose that Assumptions \ref{ass:w(z)},  and S1-S4 hold.
We have 
\begin{align*}
\|\bthetahat\supmeta- \bthetabar\|_1
= O_p &\left\{  s_\btheta \sqrt{\frac{\log~d}{n}}  +  \|\bdel_0\|_1  + \sqrt{\frac{n}{\log~d}} \rm{Err}_\bz \left(1 \vee \sqrt{\frac{n}{n\submT}}\rm{Err}_{\bz}  \right) \right\},\\
\text{and} \quad \|\bthetahat\supmeta- \bthetabar\|_2^2 = O_p &\left\{
s_\btheta \frac{\log~d}{n} + 
\sqrt{\frac{\log~d}{n\submT}}\|\bdel_0\|_1 \wedge \|\bdel_0\|_1^2 + \rm{Err}_\bz \left(1 \vee \frac{n}{n\submT} \rm{Err}_\bz\right) \right\},
\end{align*}
\end{Th}
where ${\rm Err}_\bz = n\submS^{2}n^{-2} (\|\widehat\bz - \bz_0\|_2^2 + n\submS^{-1}\log~d\|\wh \bz - \bz_0\|_1^2 )$.
\begin{Rem} 
As one referee pointed out, the $l_1$ error rate in Theorem \ref{th:1} is no better than the rate of $\|\wt \btheta - \btheta^*\|_1 + \|\btheta^* - \bar\btheta\|_1$, which implies that the second bias correction step in Algorithm \ref{alg} does not improve the $l_1$ error rate when $\bdel_0$ is $l_1$ sparse. 
This is because of the intrinsic properties of applying $l_1$ type penalties to $l_1$ sparse parameters, and it agrees with the minimax result proved in \cite{li2020transfer} and \cite{li2023estimation}. 
\cite{negahban2012unified} also shows that for high dimensional linear models, the $l_1$ error of a penalized estimator with a decomposable penalty for a $l_1$ sparse parameter is the $l_1$ norm of the parameter itself, which is also the minimax rate \citep{raskutti2011minimax}.
However, this bias correction step can help improve the $l_2$ error.
Moreover, when we consider $l_r$ sparse $\bdel_0$, $r\in [0,1)$, the rate of
$\|\widehat\btheta\supmeta - \bar\btheta\|_1$ can be better than that of $\|\wt \btheta - \btheta^*\|_1 + \|\btheta^* - \bar\btheta\|_1$. Detailed results are given in Corollaries S1 and S2 in Supplementary Materials. 
For example, when both $\bar\btheta$ and $\bdel_0$ are $l_0$ sparse and $\|\bdel_0\|_0 \precsim s_\btheta$, 
we have 
$\|\wt \btheta - \btheta^*\|_1 = O_p
\left(s_\btheta \sqrt{n^{-1}\log~d}  + \rm{Err}_\bz\sqrt{n/\log~d}\right)$
and 
$
\|\widehat\btheta\supmeta - \bar\btheta\|_1 = O_p\left\{
s_\btheta \sqrt{n^{-1}\log~d} + \|\bdel_0\|_0 \sqrt{n\submT^{-1}\log~d} + \rm{Err}_\bz\sqrt{n/\log~d} (1 \vee \sqrt{n/n\submT}{\rm Err}_\bz)\right\}.
$
When $\|\bdel_0\|_0 \sqrt{n\submT^{-1}\log~d} =o(\|\bdel_0\|_1)$, the bias correction step helps. 
\end{Rem}

In typical SS settings, we have $\max(n\submS, n\submT) = o\{\min(N\submS, N\submT)\}$, resulting higher order estimation error for $\wh\bz$, and simplified error bounds
$
\|\bthetahat\supmeta- \bthetabar\|_1
= O_p \left(  s_\btheta \sqrt{n^{-1}\log~d}  +  \|\bdel_0\|_1  \right)$, and 
$
\|\bthetahat\supmeta- \bthetabar\|_2^2 =O_p \left(
s_\btheta n^{-1}\log~d + 
\sqrt{n\submT^{-1}\log~d}\|\bdel_0\|_1 \wedge \|\bdel_0\|_1^2 \right).
$
When  $w(\z)$ or $m(\z)$ is correctly specified,  then $\bdel_0=\0$,
$\|\bthetahat\supmeta -\bthetabar\|_1 = O_p\left( s_\btheta \sqrt{n^{-1}\log~d} \right)$,
$\|\bthetahat\supmeta -\bthetabar \|_2^2 =O_p \left(  s_\btheta n^{-1}\log~d \right)$,
which are the standard high dimensional error rates with a sample size of $n=n_\mS+n_\mT$.  In other words, the source labeled data have been fully utilized to improve the estimation accuracy of $\bthetabar$.
When $\|\bdel_0\|_1 =o\left(s_\btheta\sqrt{n\submT^{-1} \log~d}\right)$, even if both of the working models are misspecified, we can still gain estimation accuracy by partially utilizing the source labeled observations.   
When $ s_\btheta\sqrt{n\submT^{-1} \log~d} = o(\|\bdel_0\|_1)$, 
the error rates are worse than those of the target-only estimator and we begin to have the negative transfer.

\begin{Rem}\label{NT_sc2}
In scenario 2, where both $\bdel_0$ and $\bar\btheta$ are $l_r$ sparse for a fixed $r\in[0,1]$, by Corollary S2 in the Supplementary Materials, we begin to have negative transfer 
when $\|\bar\btheta\|_r = o(\|\bdel_0\|_r)$. It agrees with the intuition that when the difference between the shifted source parameter and the target parameter is denser than the target parameter, the second step in Algorithm \ref{alg} using target labeled data could not fully correct the bias in $\widetilde \btheta$, and hence the resulting meta-learning estimator is worse than the target-only estimator. 
\end{Rem}

We present in Theorem \ref{th:2} the upper error bound of $\bthetahat\supSTRIFLE$, which overcomes negative transfer when $\|\bdel_0\|_1$ is large. The proof is given in Section S.7 of Supplementary Materials. 
\begin{Th}\label{th:2}
	Suppose that Assumptions \ref{ass:w(z)}, and S1-S4 hold. 
	We have 
 {\small
	\[
	\|\bthetahat\supSTRIFLE - \bthetabar \|_2^2  = O_p \left[\left\{\frac{\log~d}{n} s_\btheta+ \sqrt{\frac{\log~d}{n\submT}} \|\bdel_0\|_1 \wedge \|\bdel_0\|_1^2 +\rm{Err}_\bz \left(1 \vee \frac{n}{n\submT} \rm{Err}_\bz\right)\right\} \wedge 
	\left(  \frac{\log~d}{n\submT}s_\btheta+ n\submT^{-1/2} \right)\right].
	\]	
 }
\end{Th}

We can see that when $ n\submT^{1/2}/\log~d \precsim  s_\btheta$, 
$\|\bthetahat\supSTRIFLE - \bthetabar \|_2^2 $ is guaranteed to be no worse than that of the target-only estimator $\bthetahat\supmT$. While when $ s_\btheta = o(n\submT^{1/2}/\log~d)$, there is indeed a price to pay for the transferability detection step.
Similar results have been proved for general $l_r$ sparsity. See Theorem S2 in Supplementary Materials.

\subsection[]{Theoretical Properties of $\wh \bb\supSTRIFLE$}\label{subsection:beta}

With triply robust estimator $\bthetahat\supSTRIFLE$, we can leverage target unlabeled data to robustly improve the learning accuracy of $\bb_0$.
Similar to the SAS estimator, we allow $\bb_0$ to be much denser than $\bthetabar$ when $n\submT=o(N\submT)$.
The upper error bound of $\wh\bb\supSTRIFLE$ is established in the following Theorem with its proofs given in Section S8 of Supplementary Materials.

\begin{Th}\label{th:3}
Under Assumptions\ref{ass:w(z)}, and S1 - S5, let  $\varrho = n\submT/(n\submT + N\submT)$, and we have
\begin{align*}
\|\wh \bb\supSTRIFLE-\bb_0\|_2^2 
 = O_p &\left(
(1 - \varrho)\left[\left\{ s_\btheta \frac{\log ~d}{n}+ \sqrt{\frac{\log~d}{n\submT}} \|\bdel_0\|_1 \wedge \|\bdel_0\|_1^2 + \rm{Err}_\bz \left(1 \vee \frac{n}{n\submT} \rm{Err}_\bz\right)\right\} \right.\right. \\ & \left.\left. \wedge \left( \frac{\log ~ d}{n\submT}s_\btheta + n\submT^{-1/2}\right)\right]  + s_\bb \frac{\log~p}{N\submT + n\submT} \right).  
\end{align*}
\end{Th}
\begin{Rem}
$\widehat\bb\supSTRIFLE$ is guaranteed to be no worse than the target-only supervised estimator
when $(1 - \varrho) s_{\btheta} n\submT^{-1} \log ~ d \precsim s_{\bb} n\submT^{-1} \log~ p$ and $(1-\varrho)n\submT^{1/2}/\log~p \precsim s_\bb$ even if $N\submT \precsim n\submT$. 
These conditions are typically met in EHR applications with highly predictive surrogates, where 
$\bar\btheta$ is much sparser than $\bb_0$, and in polygenic risk models with weak but dense SNP signals.
When $s_\bb = o\{(1-\varrho)n\submT^{1/2}/\log~p\}$, the error from transferability detection becomes dominant.
\end{Rem}
\begin{table}[H]
\centering
\caption{Comparison of  squared $l_2$ error bound of estimation of $\bb_0$ among three methods in SS setting with negligible $\rm{Err}_\bz$ and $1-\varrho \asymp 1$. The common term $\bDel_\bb = s_\bb (N\submT + n\submT)^{-1}\log~ p$, $\rm{cutoff}_1 = s_\btheta  n^{-1} \sqrt{n\submT\log~d}$, and $\rm{cutoff}_2 = s_\btheta\sqrt{n\submT ^{-1}\log~d } \vee  (\log~d)^{-1/2}$.  
}
	\label{t0}
	\scalebox{0.85}{\begin{tabular}{lccc}
			\multicolumn{1}{c}{Method} & SAS & Meta & STRIFLE \\ \hline 
			$0\leq\|\bdel_0\|_1 \precsim \rm{cutoff}_1 $               
			&$\bDel_\bb  +  s_\btheta n\submT^{-1}\log ~d $    
			& $\bDel_\bb  +  s_\btheta n^{-1} \log ~d $         
			& $\bDel_\bb  + s_\btheta  n^{-1} \log ~d $         \\
			$\rm{cutoff}_1  \ll \|\bdel_0\|_1 \precsim \rm{cutoff}_2$ 
			&$\bDel_\bb  +   s_\btheta n\submT^{-1} \log ~d $     
			&  $\bDel_\bb + \sqrt{n\submT^{-1}\log ~d} \|\bdel_0\|_1$       
			&    $\bDel_\bb  +\sqrt{n\submT^{-1}\log ~d}  \|\bdel_0\|_1$        \\ 
			$\|\bdel_0\|_1 \gg \rm{cutoff}_2$              
			&$\bDel_\bb   +s_\btheta  n\submT^{-1}\log ~d $     
			&    $\bDel_\bb  + \sqrt{n\submT^{-1}\log ~d }  \|\bdel_0\|_1$     
			& $\bDel_\bb  + s_\btheta n\submT^{-1}\log ~d  + n\submT^{-1/2}$ \\ \hline     
	\end{tabular}}
\end{table}

\begin{Rem}
As summarized in Table \ref{t0}, in the SS setting with negligible $\rm{Err}_\bz$ and $1-\varrho \asymp 1$, 
$\bbhat\supmeta = \wh\bb(\wh\btheta\supmeta)$ and $\bbhat\supSTRIFLE$ both outperform the SAS estimator when $\|\bdel_0\|_1 = o(s_\btheta\sqrt{n\submT^{-1} \log~d})$ by successfully transfering useful source information.
When the model-assisted heterogeneity between the two populations is too large, i.e.,  $s_\btheta\sqrt{n\submT^{-1} \log~d} = o(\|\bdel_0\|_1)$,  $\bbhat\supmeta$ is worse than the SAS estimator due to negative transfer. 
However, the squared $l_2$ error of $\bbhat\supSTRIFLE$ is upper bounded by that of the SAS estimator plus the error from transferability detection, which is negligible when $n\submT^{1/2}/\log~d \precsim  s_\btheta$.
In conclusion, $\widehat\bb\supSTRIFLE$ is triply robust against nuisance model misspecification, shift in $p_{Y\mid\Z}$, and negative transfer. 
\end{Rem}

\section{Simulation Studies} \label{sec:simulation}

In this section, we evaluate the finite-sample performance of the proposed STRIFLE estimator for both estimation and prediction accuracy. We compare the STRIFLE estimator with four benchmark methods: (i) the supervised estimator (SUP) obtained by estimating $\bb_0$ directly using data in $\mathcal{I}\submTL$; (ii) the SAS estimator obtained by first estimating $m(\z)$ using data in $\mathcal{I}\submTL$ then leveraging data in $\mathcal{I}\submTU$ by minimizing (\ref{eq:lossbb}); (iii) the covariate shift estimator (CS) obtained by first estimating $w(\z)$ using data in $\mathcal{I}\submTU$ and $\mathcal{I}\submSU$ then estimating $\bb_0$ by minimizing 
$-{n}^{-1}\sum_{i\in \mathcal{I}_{\mS, L}\cup\mathcal{I}\submTL}   \wh w(\z_i)^{I(i\in \mathcal{I}_{\mS, L}) }
Q\left\{ g(\x_i\trans \bb), y_i\right\}  +\lambda_\bb \|\bb\|_1$;
and (iv) the supervised transfer learning estimator (SUPTrans) proposed in \cite{li2021targeting}, which utilizes the source labeled data through a three-step transferring procedure with a stronger sparsity assumption on $\bb_0$ without leveraging the unlabeled data.

Throughout, we consider binary outcomes with $g(\cdot)$ being anti-logit and use the Lasso penalty for all models for computational ease. All tuning parameters of the Lasso penalties are selected via 5-fold cross-validation.  We fix $n\submT = 150$, $N\submT = N\submS = 10000$,  $q=30$. Let $p=150$ or $1000$, and $n\submS = 150$ or $1200$. Since $N\submS$ and $N\submT$ are much larger than $d$,
we estimate $\bz\in  \mR^d$ by first minimizing $\mL_{N}^w(\bz) +\lambda_\bz\|\bz\|_2^2$ with a small $\lambda_\bz$ to stabilize the optimization, and then applying hard thresholding with a cutoff $10\sqrt{\log~d/(N\submS +N\submT)}$.  

Instead of generating data separately for $\mT$ and $\mS$, we generate $\Z_i$ and $Y_i$ for $\Dscr\submS\cup\Dscr\submT$ and then randomly generate a membership variable $R_i = I(i \in \mI\submSL\cup\mI\submSU)$ to assign the $i$th observations to the source population when $R_i = 1$. Specifically, we generate $\Z_i$ mimicking the zero-inflated discrete distribution of EHR features. We first generate latent variables  $\W_i \sim N(\0, \bSig)$, where $\bSig = [\nu^{|j-k|}]_{d\times d}$, for $j, k = 1, \dots, d$.
Then we generate  $X_{ij}  = a(W_{ij})$, for $j=1, \dots, p$,  $S_{i1} =a( \sum_{j=1}^5X_{ij} + W_{i,p+1}) $, $S_{i2} = a( \sum_{j=3}^{7}X_{ij} + W_{i,p+2})$, $S_{i3} = a( \sum_{j=6}^{10}X_{ij} + W_{i,p+3})$, and $S_{ij} = a(W_{i,p+j})$, for $j=4, \dots, q$, where the transformation $a(t) = [\log(1 + e^t)]$, and $[x]$ means to take the integer part of $x$. $\X$  and $\S$ are standardized to have mean zero and variance one. We consider $\nu = 0$ resulting $\bSig=\I_p$ and $0.1$ allowing for weak correlations among elements of $\W_i$. 
We consider two models for $Y_i \mid \Z_i$:
\begin{alignat*}{2}
	& {\rm(M\subcor)} &\quad &  \logit\{\Pr(Y_i=1\mid \Z_i)\} = 0.2  - X_{i1} + X_{i2} -X_{i3} + 2S_{i1} - 2S_{i2} +2S_{i3}; \\
	& {\rm(M\submis)}  &\quad & 	\logit\{\Pr(Y_i=1\mid \Z_i)\} = -X_{i1} + X_{i2}  + X_{i3}^2  - S_{i2} + 2S_{i3} + \frac{2S_{i1}}{1 + \exp(- X_{i1}S_{i1}^2)}.
\end{alignat*}
The imputation model $m(\z)=g(\z\trans \btheta)$ is correctly specified under (M$\subcor$) but misspecified under (M$\submis$). For the misspecified scenario under (M$\submis$),  $\bthetabar$ is soft sparse when $\nu=0.1$.
We consider two models for generating $R_i$:
\begin{alignat*}{2}
	& {\rm(W\subcor)} & \quad &  \logit\{\Pr(R_i=0 \mid \Z_i)\} =X_{i1} -X_{i2}  - X_{i3} + S_{i1}, \\
	& {\rm(W\submis)} & \quad & \logit\{\Pr(R_i=0\mid \Z_i)\} =  1.8S_{i2} - 2S_{i3} + S_{i1} \left[1 + \frac{ X_{i1} + X_{i2} }
	{1 + \exp \{- (\sum_{j=3}^{10} X_{ij})\}}\right].
\end{alignat*}
We set $\f(\z)$ to $\z$ when fitting the parametric density ratio model (\ref{model-DR}), leading to correct specification of (\ref{model-DR}) with (W$\subcor$) and misspecification with (W$\submis$).
We have four different configurations by combining (M$\subcor$), (M$\submis$), (W$\subcor$) and (W$\submis$) under two correlation structures for $\Z$, two values for $n\submS$ and two values for $p$.  

To examine the potential negative transfer consequence from different estimators and demonstrate the triple robustness of our STRIFLE estimator, we additionally consider simulation settings when both  $m(\z)$ and $w(\z)$ are severely misspecified. The theoretical analysis suggests that negative transfer can be more pronounced for the intermediate Meta estimator when $s_\btheta\sqrt{n\submT\log~d} = o(\|\bdel_0\|_1)$.  To this end, we generate $Y_i$ and $R_i$ from
{\small\begin{alignat*}{2}
&{\rm(M\submis^\prime)} & \quad &  \logit\{\Pr(Y_i=1\mid \Z_i)\} = -1-2X_{i1} + 2X_{i2}^2 + 2X_{i3}X_{i4}  + 4S_{i1}^2 - 4S_{i2}^2 + \frac{1}{1 + \exp\{- X_{i3}(S_{i3}^2 + S_{i4})\}},  \\
&{\rm(W\submis^\prime)}& \quad &  \logit\{\Pr(R_i=0\mid \Z_i)\} =  4X_{i1} - 4X_{i2} - 6S_{i1}^2 + 6S_{i2}^2 - \frac{2S_{i3}(X_{i1} + X_{i2})}{1 + \exp \{- (\sum_{j=3}^{10} S_{ij})\}}, 
\end{alignat*}}
respectively with $\nu=0$. 
The rest of the settings remain the same as above.

For each setting, we performed 500 simulation replications to assess the average performance of various estimators.
We measured the average absolute bias across $p$ components of $\bbhat$ ($|$Bias$|$), $l_2$ error ($l_2$ Err) of $\bbhat$, and the area under the ROC curve (AUC) of $\bbhat\trans\X$ for classifying $Y$. AUC was calculated using the 10000 target unlabeled observations.
We first summarize the results in Table \ref{t1} for settings where $p=1000$ and at least one of $m(\z)$ and $w(\z)$ is correctly specified, i.e., $\bdel_0 = \0$. Across all settings considered, STRIFLE generally outperforms SUP, SAS, CS and SUPTrans with respect to estimation and prediction accuracy by successfully leveraging target unlabeled data and fully utilizing source labeled data.  
STRIFLE substantially beats SAS, which is also robust against the misspecification of $m(\z)$ without employing $w(\z)$,  when $n\submS$ is much greater than $n\submT$, because more useful information can be borrowed from the source population.
STRIFLE also significantly surpasses SUPTrans when the heterogeneity between source (without shift based on $w(\z)$) and target populations is large in terms of $\bb$, $N\submT$ is much greater than $n\submS$, and/or $s_\bb$ is much denser than $s_\btheta$. 
Further, the performance of STRIFLE with $\nu=0.1$ is very similar to that with $\nu=0$, 
demonstrating that STRIFLE can also be used when $\bar\btheta$ is soft sparse.

Table \ref{t2} summarizes results for configurations 
(C4): (M$\submis$)+(W$\submis)$ and (C5): (M$\submis'$)+(W$\submis')$ with more severe model misspecification in (C5).
For illustration, we also included results from $\wh\bb\supmeta$, which is expected to suffer from negative transfer under severe model misspecifications.
Under configuration (C5), as expected, both Meta and CS estimators underperform the SAS estimator due to the negative transfer.
However, STRIFLE maintains performance on par with SAS, demonstrating adaptability to heterogeneity and resistance to negative transfer. 
On the other hand, under configuration (C4) where the two populations are reasonably similar with a smaller magnitude of $\|\bdel_0\|_1$, STRIFLE can still partially utilize source data outperforming SAS.
For SUPTrans, its performance improves when $n\submS$ increases from 150 to 1200, but still worse than that of STRIFLE under both configurations, failing to effectively use predictive surrogates and the unlabeled data.

{\small
\begin{table}[tbp]
	\singlespacing
	\centering
	\caption{ Average absolute bias ($|$Bias$|\times 10^{-2}$), average of $l_2$ error ($l_2$ Err), area under the ROC curve (AUC)  
		of five different methods
		under  three different modeling configurations: (C1): (M$\subcor$)+(W$\subcor$);  (C2): (M$\submis$)+(W$\subcor$); (C3): (M$\subcor$)+(W$\submis$) with $p=1000$, $\nu = 0$ and $0.1$ and $n\submS =150$ and $1200$.
	}
	\label{t1}
	\begin{tabular}{lccccccccc}
		\hline
		& $|$Bias$|$         & $l_2$ Err           & AUC         & $|$Bias$|$           & $l_2$ Err          & AUC    & $|$Bias$|$         & $l_2$ Err           & AUC     \\ \hline

		$\nu =0$ 		& \multicolumn{3}{c}{ (C1) Both correct}     & \multicolumn{3}{c}{(C2) $m(\z)$ misspecified}  
		& \multicolumn{3}{c}{(C3) $w(\z)$  misspecified} \\ \cline{2-10} 
		
		SUP  &0.86 &1.16 &0.72 
		& 0.99   & 1.39   &0.62  
		&0.92 &1.33 &0.61 \\
		SAS &0.81 &1.06 & 0.75 
		&0.93    & 1.29   &0.67
		&0.84 &1.16 & 0.69\\ \cline{2-10}
		CS$_{n\submS = 150}$ &0.84 &1.11 & 0.73 
		&0.96    &1.31 & 0.65
		&0.89  &1.27 &0.63 \\
		SUPTrans$_{n\submS = 150}$ &0.89 &1.23 & 0.73
		&1.01 &1.39 & 0.66
		&0.94 &1.28 & 0.64\\ 
		STRIFLE$_{n\submS = 150}$  &\bf{0.78} &\bf{0.95} &\bf{0.76}  &\bf{0.89}&\bf{1.18}&\bf{0.70} 
		&\bf{0.81} &\bf{1.10} & \bf{0.70}\\ 
		\cline{2-10} 
		CS$_{n\submS = 1200}$ &0.77 &0.84 &0.77
		&0.84 & 0.93& 0.73
		&0.85 &1.08 & 0.68\\ 
		SUPTrans$_{n\submS = 1200}$ &0.82 &0.95 & 0.79 
		&0.90 & 1.02 & \bf{0.76} 
		&0.88 &1.07 & 0.69
		\\ 				
		STRIFLE$_{n\submS = 1200}$ &\bf{0.72} &\bf{0.64}  &\bf{0.80}
		&\bf{0.79}&\bf{0.78}& 0.76
		&\bf{0.74} &\bf{0.77} & \bf{0.75}\\ 				
		\hline

		$\nu = 0.1$		& \multicolumn{3}{c}{(C1) Both correct }& \multicolumn{3}{c}{(C2) $m(\z)$  misspecified }
		&\multicolumn{3}{c}{(C3) $w(\z)$  misspecified }
		\\ \cline{2-10} 
		
		SUP   &0.87  & 1.20 &0.71   
		&1.01 &1.40 &0.63 
		&0.95 &1.38 &0.60 \\
		SAS  &0.83 &1.11 & 0.73        
		&0.95 &1.29 &0.69 
		&0.86 &1.23 & 0.69\\ \cline{2-10}
		CS$_{n\submS = 150}$ &0.85 &1.14 &0.72     
		&0.97 &1.30 &0.67
		&0.91 &1.32 &0.63
		\\
		SUPTrans$_{n\submS = 150}$  &0.91 &1.28 & 0.71   
		&1.02 &1.40  & 0.67
		&0.96 &1.34 &0.63
		\\ 
		STRIFLE$_{n\submS = 150}$  &\bf{0.80} &\bf{1.00}& \bf{0.75}    
            &\bf{0.90} &\bf{1.18} & \bf{0.71}
		& \bf{0.84}&\bf{1.16}&\bf{0.70}\\ 
		
		\cline{2-10} 
		
		CS$_{n\submS = 1200}$  &0.77 &0.84 &0.77  
		&0.85 &0.92 &0.75 
		& 0.86  &1.09 &0.69  \\
		SUPTrans$_{n\submS = 1200}$      & 0.83 &0.96 &0.79
		&0.91 &1.02 &\bf{0.77} 
		& 0.90 & 1.11 & 0.70  \\ 
		STRIFLE$_{n\submS = 1200}$ &\bf{0.73}&\bf{0.67}&\bf{0.80}  
        &\bf{0.80} &\bf{0.80} &0.77 
		&\bf{0.75}  &\bf{0.80} &\bf{0.75} 
		\\ 
		\hline
		
	\end{tabular} 
\end{table}
}

{\small
\begin{table}[tbp]
	\singlespacing	
	\centering
	\caption{ Average absolute bias ($|$Bias$|\times 10^{-2}$), average of $l_2$ error ($l_2$ Err), area under the ROC curve (AUC)  
		of six different methods
		under two configurations, where both nuisance models are misspecified with  p=1000. Negative transfer does not show up in (C4) but does exist in (C5). 
	}
	\label{t2}
	\begin{tabular}{lcccccc}
		\hline
		& $|$Bias$|$         & $l_2$ Err           & AUC            & $|$Bias$|$         & $l_2$ Err           & AUC     \\ \hline
	 $\nu = 0$	& \multicolumn{3}{c}{(C4) $(M\submis) + (W\submis)$   }     
		& \multicolumn{3}{c}{(C5) $(M\submis^\prime) + (W\submis^\prime)$} \\ \cline{2-7} 
		
		SUP &1.03  &1.51   & 0.61
		&0.94 &1.32 & 0.73\\
		SAS & 0.90 & 1.24 & 0.73
		&\bf{0.86} &\bf{1.08} & \bf{0.78}\\  \cline{2-7}
		CS$_{n\submS = 150}$ & 1.00 & 1.42 & 0.65
		&1.09  &1.91 &0.63\\
		SUPTrans$_{n\submS = 150}$ & 1.04 & 1.53 & 0.62
		&0.96 &1.35 & 0.73\\ 
		Meta $_{n\submS = 150}$  &\bf{0.86} &\bf{1.09} & \bf{0.75}
		&0.95 &1.39  &0.73\\ 
		STRIFLE$_{n\submS = 150}$ &0.87 & 1.11 & 0.74
		&0.87 &1.12  &0.77\\ 
		\cline{2-7} 
		CS$_{n\submS = 1200}$ &0.90 & 1.09 & 0.74
		&1.18 &2.68 & 0.60\\
		SUPTrans$_{n\submS = 1200}$ &0.96 & 1.26 & 0.68
		&0.94 &1.32 & 0.76\\ 	
		Meta $_{n\submS = 1200}$ &\bf{0.78} & \bf{0.75} & \bf{0.79}
		&0.97 &1.48 & 0.69\\
		STRIFLE$_{n\submS = 1200}$ & 0.78 & 0.76 & 0.79
		&0.87 &1.13 & 0.77
		\\ 				
		\hline
	\end{tabular} 
\end{table}
}

The results with $p=150$ are summarized in Tables S1 and S2 in Section S.10 of Supplementary Materials. 
We further explored the use of an NN with 10 layers based on scaled conjugate gradient backpropagation to estimate $w(\z)$ in the CS, Meta, and STRIFLE methods.
When $N\submS$ and $N\submT$ are sufficiently large, the performance using NN is comparable to that of penalized parametric estimation of $\bz$ when $w(\z) = \exp(\z\trans \bz)$ is correctly specified. While NN excels when $w(\z)$ is misspecified with a relatively simple structure.
In (C3) and (C4), NN enhances the transferability of source labeled data. 
In contrast, in (C5) where $w(\z)$ is severely misspecified with complex nonlinearity, NN and penalized parametric estimation of $\bz$ yield similar results. Meta with NN is still worse than SAS, indicating negative transfer. 
In summary, when $N\submS$ and $N\submT$ are insufficient for complex ML models, the error of density ratio estimation can result in negative transfer, making the bias correction and transferability detection steps essential. 
Additionally, we also investigated the impact of binarizing $g(\z_i\trans\widehat\btheta)$ using the target sample prevalence as the cutoff for imputation in SAS and STRIFLE methods. In our simulation studies, this approach yields slightly inferior results compared to directly using $g(\z_i\trans\widehat\btheta)$ for imputation.

\section{T2D Genetic Risk Prediction for African Americans}\label{sec:realdata}
In this section, we apply the proposed STRIFLE method to derive a T2D polygenic risk prediction model for the AA population, using SNP and EHR data from  Mass General Brigham's (MGB) Biobank. 
In this study, more than 87\% of the participants are of European descent, but only around 5\% are AA. We chose the AA group as our target population and the European ancestral group as our source population. 
The binary response variable $Y$, which represents the manual chart review label for T2D, is available in limited quantities from both populations. There are $n\submT = 95$ labeled observations and $N\submT = 2900$ unlabeled observations from the AA population; $n\submS = 1137$ labeled observations and $N\submS = 50242$ unlabeled observations from the European population.
Our aim is to develop a polygenic risk prediction model for $Y$ based on SNPs, age, and gender for the target AA population, with the aid of data from the larger European population and observations on $\S$, consisting of counts of T2D-related EHR codes selected in \cite{hong2021clinical} via a knowledge graph.
Due to potential systematic coding biases in International Classification of Diseases (ICD) billing codes across different demographic populations, we removed ICD codes and only used 76 laboratory tests and medication prescriptions as surrogates. 
Examples of surrogates include the count of HbA1c tests and the number of metformin prescriptions, 
a common treatment for controlling blood sugar in T2D patients. 
All count variables are transformed by $x \to \log(1+x)$ due to their skewness. We identified $248$ SNPs from those previously reported as predictive of T2D, low--density lipoprotein, or high--density lipoprotein cholesterol gene \citep{kozak2009ucp1,rodrigues2013genetic} with additional removal of SNPs with minor allele frequency less than 0.08 and correlations among the SNPs greater than 0.5. Together with age and gender, we have $250$ candidate predictors to form $\X$.  
We normalized $\X$ and $\S$ to have zero mean and unit variance within each population.

Similarly to the simulation studies, we derive T2D polygenic risk prediction models using STRIFLE, SUP, CS, SUPTrans and SAS methods. We employed 10-fold cross-validation for tuning parameter selection. To compare the performance of different methods, we used 10-fold cross-validation to estimate the out-of-sample AUC and mean squared error (MSE), which is defined as the mean of squared differences between the predicted probability of T2D  and the true binary T2D status.
This process was repeated 20 times, and we reported the average out-of-sample AUC and MSE for each method.
We also used source labeled data to estimate $\bb$ then validated on target labeled data to illustrate potential heterogeneity between populations, denoted by SUPSource. 
As summarized in Table \ref{t3},
the STRIFLE estimator yields the best estimation and prediction accuracy by successfully utilizing the source population and the target unlabeled data.
{\small\begin{table}[H]
	\singlespacing
	\centering
	\caption{Prediction (AUC) and Estimation (MSE) accuracy  of estimated $\bb_0$ from SUP, CS, SUPTrans, SAS and STRIFLE methods, compared to supervised source estimator (SUPSource) evaluated on target labeled data.}
	\label{t3}
	\begin{tabular}{c|c|ccccc}
		\hline
		& Source Estimate  & \multicolumn{5}{c}{Target Estimate}\\\hline
		& SUPSource & SUP     & CS     & SUPTrans  &SAS & STRIFLE \\
		\hline
		AUC  & 0.77
		&0.66 &0.77  &0.76    & 0.79 & \bf{0.85} \\
		\hline
		MSE & 0.21 
		&0.21 &0.21 &0.20  &0.19 &\bf{0.17}  \\
		\hline
	\end{tabular}
\end{table}
}
The estimates of $\bb_0$ from all methods are presented in Section S.11 of Supplementary Materials. 
Additionally, Figure \ref{fig:1} shows the coefficient estimates for the top signals with an absolute magnitude above 0.1 for each method. Consistent with expectations, age emerges as the most significant predictor across all methods, exhibiting similar estimated effect sizes. However, gender, while displaying a higher magnitude in the source population, is not identified as an informative predictor in the SUP, SAS, and STRIFLE methods.
The SNP rs7903146, in line with findings from \cite{grant2006variant}, was selected by all methods. This reflects its known association with T2D. 
In general, the SNPs chosen by SUPSource significantly differ from those selected by the other methods, highlighting the heterogeneity between populations and the distinctions between 
 $E\submS(Y\mid \X)$ and $E\submT(Y\mid \X)$.
In consequence, different risk prediction models for T2D should be employed for the European and AA ancestral groups to achieve algorithm fairness. 
For AA people, the two most significant SNPs related to T2D  based on our STRIFLE method are rs174546, which is closely related to fatty acid desaturase \citep{billings2010genetics, sergeant2012differences}, and rs7903146, which is suggested as the causal diabetes susceptibility variant in AAs in \cite{palmer2011resequencing}.

\begin{figure}
	\centering
	\caption{The estimates of $\bb_0$ for top signals with absolute magnitude above 0.1 from SUP, CS, SUPTrans, SAS, and STIFLE with SUPSource as a reference.}
	\includegraphics[width=0.6\textwidth]{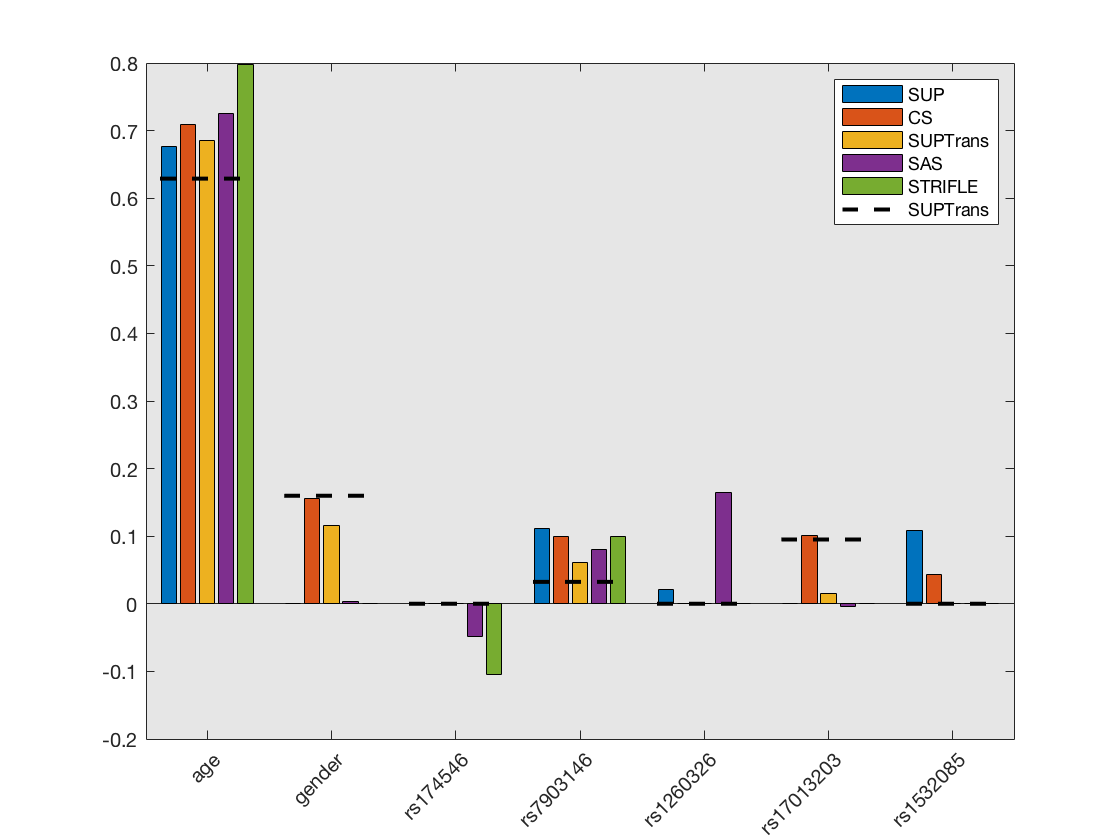}
	\label{fig:1}
\end{figure}

\section{Conclusion and Discussion}\label{sec:dicussion}
In this paper, we have proposed an adaptive transfer learning method in the SS learning framework. 
Different from other existing methods, our method is both model-assisted and surrogate-assisted. 
Compared with existing non-model-assisted transfer learning methods, 
when a large number of unlabeled observations from both populations are available, which is common in large-scale biomedical studies, the density ratio model can be estimated relatively well, which consequently can improve the transferability of the source population. 
By employing the imputation model with predictive surrogates, the target unlabeled observations can be effectively leveraged and the sparsity assumption on the final outcome model can be relaxed, which is important in practice. For example, in polygenic risk prediction models, usually the signals of SNPs are relatively weak and dense. 
Instead of combining these two strategies scrappily, we construct our method, specifically the estimating equations, carefully so that it is robust against severe model misspecifications,  distributional assumption violation, and negative transfer.

The proposed triply robust transferring procedure focuses on the estimation for the imputation model under the covariate shift setting.
In fact, the source unlabeled observations can only be transferred to improve the estimation rate of $\bb$ when the density ratio model is correctly specified, since $p_{Y\mid \X}\supmT$ can be different from $p_{Y\mid \X}^{(\mS)}$.
How to robustly and efficiently conduct transfer learning directly on the outcome model, i.e., on $\bb$, in the SS learning framework is worth further research and beyond the scope of this paper.
Moreover, the theoretical results in Theorems \ref{th:1} and \ref{th:2} shed light on
transfer learning of other high dimensional nonconvex problems under the restricted strong convexity condition.

\subsection*{Acknowledgement}
The authors thank the editor, the associate editor, and the three referees for their constructive comments and suggestions, which have led to significant improvements in the article. This research was supported by NIH grants R01LM013614 and R01HL089778, and by the Center for Health and Business at Bentley University.

\subsection*{Disclosure Statement}
The authors report there are no competing interests to declare.

\bibliographystyle{agsm}
\bibliography{TransL}

\end{document}